\documentclass[a4paper,11pt]{article}

\usepackage{mathtools,mathrsfs,amssymb,amsfonts,graphicx,setspace}

\usepackage{fullpage}

\usepackage{booktabs} 
\usepackage{array}    
\newcolumntype{L}{>{$}l<{$}} 
\newcolumntype{R}{>{$}r<{$}} 
\newcolumntype{C}{>{$}c<{$}} 

\usepackage{cite} 

\usepackage{hyperref} 

\title{New (maximal) gauged supergravities}
\author{Gianluca Inverso, Mario Trigiante}

\begin{document}

\vspace*{10mm}

\begin{center}
  {\LARGE \sc 
New (maximal) gauged supergravities
   }
    \\[13mm]

{\large
Gianluca Inverso${}^{1}$ and Mario Trigiante${}^{2,3}$}

\vspace{8mm}
${}^1${\it INFN, Sezione di Padova \\
    Via Marzolo 8, 35131 Padova, Italy}
\vskip 1.2 ex
${}^2${\it Politecnico di Torino, Dipartimento di Scienza Applicata e Tecnologia,\\ corso Duca degli
Abruzzi 24, 10129 Torino, Italy}
\vskip 1.2 ex
${}^3${\it INFN, Sezione di Torino\\ Via P. Giuria 1, 10125 Torino, Italy}

\end{center}

\vspace{5mm}

\begin{center} 
\hrule

\vspace{6mm}

\begin{tabular}{p{14cm}}
{\small%
We review recent progress in constructing maximal, classical supergravity models and their applications.
\vskip2ex
\textit{%
Contribution to the book ``Half a century of supergravity'', eds.~A.~Ceresole and G.~Dall'Agata}
}
\end{tabular}
\vspace{5mm}
\hrule
\end{center}

\vspace*{10mm}

\section{Aspects of extended supergravities}

Extended supergravity theories are often encountered in the context of holography, of black hole physics in quantum gravity, and flux compactifications in string/M-theory.
This is so because under certain conditions they can capture a sector of the low-energy dynamics of string theory.
Extended supergravities (featuring 8 supercharges or more, i.e. $\mathcal{N}\ge 2$ in $D=4$) are interesting in their own right,
as they exhibit a particularly rich mathematical structure {which follows from the fact that scalar fields enter the same supermultiplets as $p$-forms (i.e. rank-$p$ antisymmetric tensor fields such as, for $p=1$, vector fields in four dimensions). The scalar fields $\phi\equiv (\phi^s)$, $s=1,\dots, n_s$,  are described, in supergravity, by a non-linear sigma model on a Riemannian manifold $\mathscr{M}_{scal.}$ (the scalar manifold).}
In the \emph{ungauged} versions of extended supergravities, where no field is minimally coupled to the vectors, supersymmetry forbids the presence of a scalar potential and mass terms.
The same symmetry implies the existence of a flat vector bundle structure on 
the scalar manifold which fixes the non-minimal coupling of the scalar fields to their $p$-form superpartners, described by sections of these vector bundles. For instance, in four dimensions, the non-minimal coupling of the scalar fields in the vector multiplets to the vector fields is described by the special K\"ahler geometry on the corresponding scalar manifold which features a symplectic vector bundle structure. 
The group $\mathcal G$ of isometries of the scalar manifold, which are global symmetries of the sigma-model Lagrangian, has a natural constant linear action on the $p$-forms.\footnote{The property of this linear transformation to be described by a constant matrix is related to the flatness of the vector bundle.} 
For instance, for $D\neq 2(p+1)$, the general form of the kinetic terms for the $p$-forms is proportional to $\mathcal{M}_{MN}(\phi)\, \mathcal{F}^M_{(p+1)}\wedge {}^*\mathcal{F}^N_{(p+1)}$,
where ${}^*$ denotes Hodge duality, $\mathcal{F}^M_{(p+1)}$ are the field strengths of the $p$-form fields, and $\mathcal{M}_{MN}(\phi)=\mathcal{M}_{NM}(\phi)>0$ encodes the non-minimal coupling to scalars, with $M=1,\dots,n_p$.
{The $p$-forms transform in a linear representation ${\bf R}_p$ of $\mathcal G$, in terms of constant matrices, while $\mathcal{M}_{MN}(\phi)$ transforms, under the action of the same group on the scalars, as a metric on the $n_p$-dimensional fiber, described by the $p$-forms themselves. As a consequence of this, 
the combined action of $g\in\mathcal G$ on the scalars and $p$-forms
extends} to a symmetry not only of the kinetic terms, but of the full set of equations of motion and Bianchi identities.%
\footnote{When quantum corrections are taken into account, the global symmetry group $\mathcal{G}$ is broken to a discrete subgroup $\mathcal{G}(\mathbb{Z})$ which preserves the lattice of quantized $p$-form charges.}

In even dimensions $D=2k$, the $n_p$ rank-$p=(k-1)$ antisymmetric tensor fields and their magnetic duals, whose $k$-form field strengths are related by Hodge duality, are both described within the same fibre of the vector bundle, whose real dimension is then~$2n_p$. 
In this case, the structure group of the vector bundle reduces to ${\rm Sp}(2n_p,\mathbb{R})$ for odd $p$ and ${\rm O}(n_p,n_p)$ for even $p$. 
The $2n_p$-dimensional linear representation ${\bf R}_p$ of $\mathcal G$ preserves this structure.
One then has that $p$-forms in $D=2(p+1)$ dimensions satisfy the $\mathcal G$ invariant twisted self-duality condition \cite{Cremmer:1979up}\footnote{We use indexless notation with $\boldsymbol{\mathcal{M}}(\phi)\equiv \big(\mathcal{M}_{MN}(\phi)\big)$, $\boldsymbol{ \mathcal{F}}_{(p+1)}\equiv \big(\mathcal{F}^M_{(p+1)}\big)$.}
\begin{equation}\label{tsd}
 {}^* \boldsymbol{ \mathcal{F}}_{(p+1)} =-{\boldsymbol\Omega}\cdot {\boldsymbol{\mathcal{M}}}(\phi)\cdot \boldsymbol{ \mathcal{F}}_{(p+1)}\,,
\end{equation}
where $\boldsymbol\Omega$ is the symplectic/pseudo-orthogonal invariant.
This equation reduces the number of propagating degrees of freedom described by the $2n_p$ fields strengths, to those of $n_p$ rank-$p$ antisymmetric tensors.
These remarkable relations between $p$-form dualities and symmetries of the scalar manifold in extended, ungauged supergravities were first observed in $D=4$ by Gaillard and Zumino~\cite{Gaillard:1981rj}.

In theories where the scalar manifold is a symmetric space, i.e. it is of the form:
$$\mathscr{M}_{scal.}=\frac{\mathcal{G}}{K(\mathcal G)}\,,$$
where $\mathcal{G}$ is a non-compact, semisimple Lie group and $K(\mathcal G)$ is its maximal compact subgroup, the above structures are better described by introducing a coset representative $\mathcal V(\phi)$, such that, for any $g\in \mathcal{G}$:
\begin{equation}\label{coset trf}
  g\cdot \mathcal{V}(\phi)=\mathcal{V}(\phi')\cdot h(g,\phi)\,,
\end{equation}
where $h(g,\phi)\in K(\mathcal G)$, and the matrix $\boldsymbol{\mathcal{M}}$ encoding the non-minimal couplings of the scalar fields to the $p$-forms has the following form:\footnote{We assume a choice of basis in the $\mathbf R_p$ module such that $\mathbf R_p[h]\cdot \mathbf R_p[h]^T = 1$.}
\begin{equation}\label{cM definition}
\boldsymbol{\mathcal{M}}(\phi)={\bf R}_p[\mathcal{V}(\phi)]\cdot {\bf R}_p[\mathcal{V}(\phi)]^T\,,
\end{equation}
which is thus fixed modulo a choice of basis of ${\bf R}_p$.

For the global $\mathcal G$ symmetry to extend to the fermionic fields (i.e. the spin-$3/2$ gravitinos and the spin-$1/2$ fields in four dimensions), 
the latter must transform in some appropriate linear representation under (the double cover of) the local $K(\mathcal G)$ through the compensating elements $h(g,\phi)$ introduced in~\eqref{coset trf}.
This fact is central to the construction of gauged supergravities.

An ungauged extended supergravity model in $D$ dimensions may  occur in off-shell inequivalent formulations, which are mapped into one another by dualizing $p$-form fields into $(D-p-2)$-forms \cite{Cremmer:1997ct}. 
The global symmetry group $\mathcal{G}$ introduced above is manifest {and maximal} when all form fields are dualized to lower-order ones so that, in particular, the scalar sector is maximal. 
In $D=2(p+1)$ dimensions one must further determine which, among the $2n_p$ $p$-forms and their duals, are to be characterized as electric fields and appear in the Lagrangian.
The global symmetry group realized locally on the Lagrangian is then a subgroup $\mathcal G_e\subset \mathcal G$ which preserves this choice. In $D=4$, being ${\bf R}_1$ symplectic, this amounts to choosing a \emph{symplectic frame}~\cite{Gaillard:1981rj}.
Ungauged maximal supergravities in $D$-dimensions are obtained from the dimensional reduction of either $11$-dimensional supergravity or ten-dimensional Type II supergravity on a torus. In their formulation with maximal $\mathcal{G}$, this group is listed in Table \ref{tab:groups and irreps}. Quantum corrections are expected to break the global symmetry group $\mathcal{G}$ to a suitable discrete subgroup $\mathcal{G}(\mathbb{Z})$ which leaves the lattice of $p$-form charges invariant. This group, encoding known string dualities, was conjectured, in the maximal four-dimensional theory, {to be an exact symmetry (U-duality) of the, as yet unknown, unique quantum theory of the fundamental interactions, unifying the known superstring models }\cite{Hull:1994ys}.
{We also refer to \cite{Nicolai:2024hqh} in this volume for further details on these models and on their exceptional duality symmetries.}

\section{Extended supergravities and their gaugings}

Although particularly appealing because of their rich global symmetry structure, ungauged extended supergravities are not phenomenologically interesting since their supersymmetry, as mentioned above, does not allow the presence either of a scalar potential or of mass terms in the Lagrangian. 
Ungauged models exhibit a continuum of Minkowski vacua parametrized by the v.e.v.s of the scalar fields, which affect the effective couplings and whose values are not restricted by any dynamics. Moreover, no spontaneous supersymmetry breaking can occur and the presence of gravity-coupled massless scalars, with the associated long-range interactions, poses serious phenomenological problems. Despite these apparent drawbacks, ungauged models have provided a useful supersymmetric field theoretical framework for the study of asymptotically flat (BPS) black hole and black brane solutions, classified by their properties with respect to the conjectured U-duality group $\mathcal{G}(\mathbb{Z})$, as well as for amplitude computations. \par

The presence of non-trivial dynamics for the scalar fields, encoded in a scalar potential, as well as of mass terms, requires the introduction of gauge interactions.
The resulting \emph{gauged} supergravities are obtained from ungauged models with the same amount of supersymmetry and field content, through the so-called \emph{gauging procedure}, which consists in promoting a suitable global symmetry group $G_g$ of the action, thus a subgroup of $\mathcal{G}$, to local symmetry. 
The gauging procedure manifestly breaks the classical global symmetry group $\mathcal{G}$ through the introduction of minimal couplings:
\begin{equation}
\mathcal{D}_\mu=\partial_\mu-g \,A^M_\mu\,X_M\,,
\end{equation}
where {the indices $M,N,\dots$ label the ${\mathbf{R}_{1}}$ representation for the vector fields and } $X_M$ are the infinitesimal generators of $G_g$.
The \emph{embedding tensor formalism}, was introduced and developed in \cite{Cordaro:1998tx,Nicolai:2000sc,deWit:2002vt,deWit:2004nw}
(see \cite{Samtleben:2008pe} for reviews), in order to describe the most general gauging of a supergravity theory in a formally $\mathcal G$-covariant fashion.
The idea is to encode all the gauge couplings in a single, non-dynamical tensor denoted by $\Theta_M{}^\alpha$, which describes the local inclusion of $G_g$ inside $\mathcal{G}$ by expressing the generators $X_M$ of the former as linear combinations of those of the latter ($t_\alpha$):
\begin{equation}
X_M=\Theta_M{}^\alpha\,t_\alpha\,\,,\,\,\,\,\,\,\alpha=1,\dots, {\rm dim}(\mathcal{G})\,.
\end{equation}
The embedding tensor {thus formally} belongs to the tensor product ${\mathbf{R}_{1}}'\times {\rm  Adj}(\mathcal{G})$ of $\mathcal{G}$-representations.
{Gauging a supergravity} amounts to {making a choice for} $\Theta_M{}^\alpha$, which therefore breaks the global symmetries $\mathcal G$ of the ungauged model.
However, the power of this formalism is that all equations of motion and Bianchi identities remain $\mathcal G$-invariant if we let $\Theta_M{}^\alpha$ transform together with the dynamical fields.
This \emph{formal} on-shell invariance should not be regarded as a symmetry of the theory, but rather as a physical equivalence between two gauged models. 
Inequivalent gauged supergravities are classified by \emph{$\mathcal G$-orbits of $\Theta$ in ${\mathbf{R}_{\Theta}}$}, modulo some consistency constraints which we now describe. 

Supersymmetry requires $\Theta$ to belong only to a certain subrepresentation ${\bf R}_\Theta\subset{\mathbf{R}_{1}}'\times {\rm  Adj}(\mathcal{G})$, which we list for maximal supergravities in table~\ref{tab:groups and irreps}.\footnote{We will mainly focus on Lagrangian gaugings, i.e. we exclude gaugings involving the on-shell $\mathbb R^+$ trombone symmetry which non-trivially acts on the Einstein frame metric~\cite{LeDiffon:2008sh}.} 
\begin{table}\centering
  \begin{tabular}{CLCC}
  \toprule
  D & \text{Global sym. group $\mathcal G$} & {\mathbf{R}_{1}}& {\mathbf{R}_{\Theta}} \\
  \midrule
  9 & \mathrm{SL}(2)\times\mathbb{R}^+ & \mathbf2_3 + \mathbf 1_{-4} & \mathbf2_{-3} + \mathbf 3_{4}  \\
  8 & \mathrm{SL}(3)\times \mathrm{SL}(2) & (\mathbf2,\,\mathbf3') 
                         & (\mathbf2,\,\mathbf3+\mathbf6')\\
  7 & \mathrm{SL}(5) & \bf10' & \mathbf{15}'+\mathbf{40} \\
  6 & \mathrm{SO}(5,5) & \mathbf{16}_c &\mathbf{144}_s \\
  5 & \mathrm{E}_{6(6)} & \bf27' & \bf351 \\
  4 & \mathrm{E}_{7(7)} & \bf56 &\bf912 \\ 
  3 & \mathrm{E}_{8(8)} & \bf248 &{\bf1}+{\bf3875} \\ 
  2 & \mathrm{E}_{9(9)} & \mathrm{basic} & \mathrm{conjugate\ basic} \\ 
  \bottomrule
  \end{tabular}
  \caption{Summary of global symmetry groups and some relevant representations for maximal supergravities~\cite{deWit:2008ta}.  \label{tab:groups and irreps}}
  \end{table}
Besides this linear constraint, consistency of the gauging requires $\Theta$ to satisfy $\mathcal{G}$-covariant quadratic constraints as well. The first such conditions follows from the requirement that $\Theta$ must select a subset of generators of $\mathcal G$ which close into the Lie algebra of a group $G_g$, and that at the same time, the vector fields involved in the gauge connection must transform in the co-adjoint representation of the same group. 
This \emph{closure constraint} reads:
\begin{equation}
    [X_M,\,X_N]=-X_{MN}{}^P\,X_P\,,\label{qc1}
\end{equation}
where $X_{MN}{}^P\equiv \Theta_M{}^\alpha\,t_{\alpha\,N}{}^P$. Condition \eqref{qc1} also implies that $\Theta$ is invariant with respect to the gauge group $G_g$ that it defines. 
In $D=4$ we also have the {\emph{locality constraint}
$
   \Omega^{MN}\Theta_M{}^\alpha\Theta_N{}^\beta=0\,,\label{qc2}
$
that ensures that} no more than $n_1$ vector fields, among the $2n_1$ electric and magnetic ones, are involved in the minimal couplings.  
\footnote{In other words, this constraint ensures that a symplectic frame exists such that $G_g\subset \mathcal{G}_e$, i.e. $G_g$ is a global symmetry of the corresponding action.}
In general, it can be shown that {this latter requirement }follows from \eqref{qc1} in theories where ${\bf R}_1$ is faithful (e.g. for $\mathcal{N}>2$) and, in the $\mathcal{N}=8$ theory, they are equivalent.

Although the gauged model is constructed from the version of the ungauged one in which all $p$-forms are dualized to lower-order ones and a maximal $\mathcal{G}$ is manifest, the full $\mathcal{G}$-covariant gauged theory requires the re-introduction of higher $p$-form fields, together with their tensor-gauge symmetries, in order to construct gauge-covariant, non-abelian field strengths.
The resulting $\mathcal{G}$-covariant couplings among forms of different orders, which are completely fixed by the embedding tensor, define the so-called \emph{tensor hierarchy} \cite{deWit:2008ta}. The resulting description of the physical degrees of freedom in terms of fields is typically redundant, as a consequence of requiring manifest covariance under $\mathcal{G}$. The embedding tensor itself fits the tensor hierarchy picture as dual to the field strengths of the non-dynamical $p=(D-1)$-forms in $D$-dimensions.

Besides gauge-covariant derivatives $\mathcal{D}_\mu$ and non-abelian gauge interactions, the gauging procedure requires the introduction of Yukawa couplings and a scalar potential in the Lagrangian, at orders $O(\Theta)$ and $O(\Theta^2)$, respectively.
These are required in order to restore the $\mathcal{N}$-extended supersymmetry that was present in the ungauged model.
Such terms are fully determined from a modification of the fermion supersymmetry variations by the addition of extra $O(\Theta)$ \emph{fermion shift terms}, which are expressed as $K(\mathcal G)$-covariant, scalar-dependent objects $\mathbb{S}_{ij}(\phi)$, $\mathbb{N}_{\mathcal{I}}{}^i(\phi)$ such that
\begin{align}
    \delta\psi_{\mu\,i}&=\mathcal{D}_\mu \epsilon_i+i\,g\,\gamma_\mu\,\mathbb{S}_{ij}(\phi)\,\epsilon^j+\dots\,,\nonumber\\
    \delta\lambda_{\mathcal{I}}&=g\,\mathbb{N}_{\mathcal{I}}{}^i(\phi)\,\epsilon_i+\dots\,,\label{fermionshifts}
\end{align}
where $\psi_{\mu\,i}$ are the $\mathcal{N}$ gravitino fields, $\lambda_{\mathcal{I}}$ are the spin-$1/2$ fields, and $\epsilon_i(x)$ are local supersymmetry parameters.
On a given vacuum,
with $\Lambda\le 0$, supersymmetry may {be (partially) }preserved depending on the values of the tensors $\mathbb{S}_{ij}(\phi_0)\,,\mathbb{N}_{\mathcal{I}}{}^i(\phi_0)$ at the extremum.

{The fermion shifts correspond to the $K(\mathcal G)$-irreducible components of the so-called \emph{T-tensor} $\mathbb{T}$ defined as the dressing of $\Theta$ by the scalar field coset representative:}
\begin{equation}
    \mathbb{T}(\phi,\Theta)\equiv\, {\bf R}_\Theta[\mathcal V(\phi)]^{-1}\cdot \Theta\,.
\end{equation}
{\sloppy{}%
In maximal $D=4$ supergravity, for instance, $\mathcal G=\mathrm E_{7(7)}$ and the double cover of $K(\mathrm E_{7(7)})$ is SU(8)~\cite{Cremmer:1979up}.
The spin 1/2 fields are denoted $\lambda_{ijk}=\lambda_{[ijk]}$ in the $\bf56$ of this group and the fermion-shift tensors\footnote{In the standard conventions for $\mathcal{N}=8,\,D=4$ supergravity, the following notation is used for the fermion-shift tensors: $A_{1\,ij}=-\sqrt{2}\,\mathbb{S}_{ij}$ and $A_{2}{}^i{}_{jk\ell}=-\frac{1}{\sqrt{2}}\,\mathbb{N}_{jk\ell}{}^i$, and the potential reads: $V(\phi)=g^2\left(\frac{1}{24}|A_{2}{}^i{}_{jk\ell}|^2-\frac{3}{4}\, |A_{1\,ij}|^2\right)$.} $\mathbb{S}_{ij},\,\mathbb{N}_{jk\ell}{}^i$ transform in the ${\bf 36}$ and the ${\bf 420}$, respectively. These representations, together with their conjugates, reproduce the branching of ${\bf R}_\Theta={\bf 912}$ with respect to ${\rm SU}(8)$:
\begin{equation}{\bf 912}\,\stackrel{\mbox{{\tiny ${\rm SU}(8)$}}}{\longrightarrow}\,{\bf 36}\oplus {\bf 420}\,\oplus \overline{\bf 36}\oplus \overline{\bf 420}\,.\end{equation}
}

To summarize, gauged extended supergravities are defined by their amount of supersymmetry, field content and local internal symmetry, encoded in  $\Theta$. They can be grouped into classes of equivalent theories whose embedding tensors are related by transformations in $\mathcal{G}$, such classes being mathematically characterized by orbits of ${\bf R}_\Theta$ with respect to the action of $\mathcal{G}$, solving the quadratic constraint~\eqref{qc1}. Full classifications of inequivalent gaugings are generally hard to achieve due to the large dimension of the ${\mathbf{R}_{\Theta}}$ representations. 
Such classification is lacking already for $D=7$ maximal supergravity (see \cite{Fernandez-Melgarejo:2011nso} for the $D=9,8$ cases).

\section{Gauged supergravities and flux compactifications}

Gauged supergravities are essential tools in
many scenarios involving flux compactifications of superstring theory and M-theory (see \cite{Grana:2005jc} for a review).
Their structure and physical properties
are 
central aspects of many applications
of the AdS/CFT correspondence \cite{Maldacena:1997re} and in the study of black-hole and brane physics, 
for instance by constructing and analyzing new anti-de Sitter vacua, domain wall solutions interpolating between them and capturing RG flows in the dual field theory \cite{Freedman:1999gp}, as well as black holes with AdS asymptotics {(see \cite{Ortin:2024slu} in this volume for a review on black hole solutions in supergravity)}.  
More specifically, gauged supergravities can encode non-trivial ten- or eleven-dimensional solutions whose spacetime geometry has the general form of a  (warped) product $M_{D}\times M_{int.}$ of a non-compact, Lorentzian $D$-dimensional spacetime $M_{D}$ and an internal compact manifold $M_{int.}$. 
Although gauged extended models can rarely be regarded as low-energy effective descriptions of string/M-theory on such backgrounds, they can arise as \emph{consistent truncations} of ten- or eleven-dimensional supergravities, which are the proper effective theories. 
Stating that a $D$-dimensional supergravity is a consistent truncation of an higher-dimensional supergravity amounts to identifying an ansatz to factorize out the dependence on the internal coordinates from all the higher-dimensional fields (and gauge parameters), in such a way that the dynamics, restricted to the factorized field space, reproduce the equations of motion of the lower-dimensional supergravity, and that solving the latter automatically solves {those} of the parent ten or eleven-dimensional model.
These scenarios generalize the standard notion of Kaluza--Klein reduction on a circle or torus to a generically curved internal manifold $M_{int.}$.
The corresponding gauged supergravity in $D$ dimensions features $M_D$ as a solution (e.g. as an (anti-) de Sitter or Minkowski maximally symmetric vacuum defined by an extremum of $V(\phi)$),%
\footnote{By vacuum solution we shall mean a maximally symmetric spacetime on which all vector and fermionic fields are zero.
} while the background quantities characterizing the internal geometry $M_{int.}$ as well as fluxes of field strengths of $p$-form fields, are all encoded in the embedding tensor $\Theta$ and thus define the gauge group $G_g$.
As such, gauged supergravities arising from consistent truncations provide an invaluable window into the non-perturbative properties of superstring theories since they capture the full non-linear dynamics of a subset of their low-lying modes on certain backgrounds.

{Perhaps the simplest examples of consistent Kaluza--Klein truncations beyond tori was devised long ago by Scherk and Schwarz~\cite{Scherk:1979zr}, who described a compactification of eleven-dimensional supergravity on a twisted version of a torus, where some cycles are non-trivially fibered over a base $S^1$. These simple geometries give rise to a maximal $D=4$ supergravity where matter fields are charged under a certain U(1) gauge symmetry and acquire masses, also giving rise to partial or total supersymmetry breaking.
This basic idea can be generalized to reductions on any internal space which is locally a Lie group manifold $G$, by factorizing the internal coordinate dependence of all fields in terms of left-invariant forms.
The supergravity models obtained from Scherk--Schwarz (SS) reductions have gauge group $G_g = G$.
Another simple setup was devised by Cremmer, Scherk and Schwarz (CSS) \cite{Cremmer:1979uq}.
In this case, one reduces a $(D{+}1)$-dimensional theory on a circle, and twists the fields along such circle by an element of the global symmetry group in $D+1$ dimensions.
Choosing a compact twist leads to a class of gaugings which generalize those obtained in the original SS paper -- albeit having only a direct $(D{+}1)$-dimensional interpretation (we come back to the $D=11$ uplift below) \cite{Andrianopoli:2002mf},\cite{deWit:2002vt},\cite{Andrianopoli:2004xu}.
}

The $\mathcal{G}$-covariant formulation of gauged supergravities in terms of the embedding tensor allows to capture the action of string dualities, encoded in $\mathcal{G}(\mathbb{Z})$, on certain types of toroidal backgrounds.
If one interprets the components of $\Theta_M{}^\alpha$ as arising from compactifications on an internal torus or group manifold, one quickly realizes that the most general embedding tensor contains, besides form-fluxes and couplings induced by SS reductions, other background quantities with no direct string/M-theory interpretation (\emph{non-geometric fluxes}).
The latter, when uplifted to the ten or eleven-dimensional parent theories, define so-called \emph{non-geometric backgrounds}, like \emph{S-folds}, \emph{T-folds} or, in general, \emph{U-folds} \cite{Hull:2003kr}, which are putative solutions to superstring or M-theory that need more than one spacetime coordinate patch to be described and such that, on overlapping patches, the transition functions also involve $T,\,S$ or, in general, $U$-dualities, within $\mathcal{G}(\mathbb{Z})$ or extensions thereof.
{Furthermore, other so-called \emph{locally non-geometric} backgrounds do not even admit a local description in terms of supergravity fields defined on coordinate patches of an underlying topological manifold.
}
It is however essential to stress that the interpretation of embedding tensor components as arising from geometric or non-geometric backgrounds is entirely dependent on the assumed topology of an underlying internal space.
What looks non-geometric on a torus may admit a perfectly geometric interpretation on, for instance, an $n$-sphere.

The most renowned examples of backgrounds captured by gauged supergravities and relevant for holography are indeed the maximally supersymmetric Freund-Rubin solutions ${\rm AdS}_4\times S^7$, ${\rm AdS}_7\times S^4$ of $D=11$ supergravity \cite{Freund:1980xh}, which describe the near-horizon geometry of stacks of M2 and M5 branes, respectively. The anti-de Sitter factors ${\rm AdS}_D$, $D=4, 7$, are solutions to maximal $D$-dimensional supergravity with gauge group $G_g={\rm SO}(12-D)={\rm Isom}(S^{11-D})$. These models describe the consistent truncations of $D=11$ supergravity to the massless anti-de Sitter supermultiplet in the Kaluza-Klein reductions on $S^{11-D}$ ($D=4,7$).  
In particular, the ${\rm SO}(8)$-gauged $\mathcal{N}=8,\,D=4$ model was first constructed in \cite{deWit:1982bul} and was proven to be a consistent truncation of eleven-dimensional
supergravity on $S^7$ in \cite{deWit:1986oxb}. Another well-known example is the ${\rm AdS}_5\times S^5$ solution to Type IIB superstring theory \cite{Schwarz:1983wa}, which describes the near-horizon geometry of a stack of D3 branes and on which the AdS/CFT correspondence was originally formulated \cite{Maldacena:1997re}. The massless gravity supermultiplet in the Kaluza-Klein compactification of the ten-dimensional theory on $S^5$ is described by the maximal $D=5$ supergravity with gauge group $G_g={\rm SO}(6)={\rm Isom}(S^5)$ \cite{Gunaydin:1984qu}. Consistency of this truncation was only recently proven in \cite{Ciceri:2014wya}. 

{
Constructing consistent Kaluza--Klein truncations on spaces such as $n$-spheres has proven significantly more complicated than in the SS and CSS examples presented above. 
The original proof of the consistent truncation of eleven-dimensional supergravity on $S^7$ relied on a highly non-trivial rewriting of eleven-dimensional supergravity in an SU(8) covariant manner. 
This approach can be regarded as a precursor of the modern frameworks  
of \emph{generalised} Scherk--Schwarz reductions in exceptional generalised geometry and exceptional field theory, {which we shall touch upon below}.
We refer to the contribution by Samtleben in this volume for more details on these frameworks.
}

In the context of maximal supergravities, \emph{exceptional generalized geometry} (EGG) ~\cite{Coimbra:2011ky} and \emph{exceptional field theory} (ExFT)~\cite{Berman:2012vc,Hohm:2013pua} have provided an ideal framework for constructing {globally and locally geometric} solutions, of the general form $M_D\times M_{int.}$, to ten-dimensional Type II or $D=11$ supergravities by uplifting to the latter models not just the corresponding $D$-dimensional vacua $M_D$, but the whole associated (maximal) gauged supergravity.
These frameworks provide a reformulation of maximal supergravities in $10$ or $11$-dimensions in which the global symmetry $\mathcal{G}={\rm E}_{11-D(11-D)}$ of a $D$-dimensional maximal model is manifest. In a nutshell, this is achieved, in both settings, by generalizing the geometry of $M_{int.}$ through an extension of the corresponding tangent space so as to accommodate the representation ${\bf R}_1$ of $\mathcal{G}={\rm E}_{11-D(11-D)}$. The latter then becomes the structure group of a generalized {tangent} bundle on $M_{int.}$, whose global structure encodes not only the internal geometry but also the $p$-form fluxes along $M_{int}$. 
{In EGG and ExFT, consistent Kaluza--Klein truncations to gauged maximal supergravities are encoded in the notion of \emph{generalised parallelizability} of such bundle~\cite{Lee:2014mla}, which is a necessary condition for the internal geometry and fluxes to support a maximal number of supercharges.
}
The factorization of the dependence on the internal coordinates of all fields and gauge parameters is encoded in a twist-matrix which is now allowed to take values in the full duality group $\mathcal{G} = {\rm E}_{11-D(11-D)}$ (times an $\mathbb R^+$ scaling factor).
{These \emph{generalized Scherk-Schwarz reductions} (gSS) include and extend the SS and CSS ones discussed earlier.}

\section{New gaugings and vacua in maximal supergravity}

Recently there has been renewed interest in classifying gauged supergravities, especially for maximal and half-maximal theories.
This is due to the combination of two technical advancements.
First, it was observed~\cite{InversoThesis,Dibitetto:2011gm,DallAgata:2011aa} that one can effectively combine the classification of gaugings with a search for maximally symmetric vacua, rephrasing the extremization conditions on the scalar potential of gauged supergravities in terms of quadratic algebraic constraints on the embedding tensor. 
{Second, the aforementioned development of EGG and ExFT allowed to clarify the structure of consistent Kaluza--Klein truncations to gauged supergravities.}

{Let us first illustrate the new approach to the search for vacua in a gauged extended supergravity. 
We focus, for concreteness, on the maximally supersymmetric model}, but the procedure works whenever the scalar manifold is homogeneous.
The scalar potential is quadratic in the embedding tensor and takes the form
\begin{equation}\label{vpot}
V\big(\phi\,,\,\Theta\big)\ =\ 
c_1\,\mathcal{M}(\phi)^{MN} \Theta_M{}^\alpha\Theta_{N}{}^\beta \big(\mathcal M(\phi)_{\alpha\beta}-c_2\,\eta_{\alpha\beta}\,\big)
\end{equation}
{where the coefficients $c_{1,2}$ depend on the spacetime dimension $D$, $\mathcal M(\phi)_{\alpha\beta}$ denotes the matrix \eqref{cM definition} in the adjoint representation of ${\rm E}_{n(n)}$, and $\eta_{\alpha\beta}$ is the $E_{n(n)}$ Cartan--Killing invariant}
{This expression holds for $D\ge4$, but the structure in $D=3,2$ is similar~\cite{Nicolai:2000sc,Eloy:2023zzh,Ortiz:2012ib,Bossard:2023wgg} (see also below).}

{When looking for vacua, solving the equations of motion reduces to extremizing the scalar potential.}
Since the scalar manifold is homogeneous, for any solution ${\phi_*}$ there exists an element $g_*\in \mathrm E_{n(n)}$ which {maps} the point ${\phi_*}$ to the `origin', which we shall denote by $\phi=0$ for simplicity, where $\mathcal M(0)_{MN}=\delta_{MN}$ and $\mathcal M(0)_{\alpha\beta}=\delta_{\alpha\beta}$.
We can therefore always write
\begin{equation}\label{gtto}
V\big(\phi\,,\,\Theta\big)\ =\ 
V\big(0\,,\,\Theta^*\big)\ =\ 
c_1\,\delta^{MN} \Theta^*_M{}^\alpha\Theta^*_{N}{}^\beta \big(\delta_{\alpha\beta}-c_2\,\eta_{\alpha\beta}\,\big)\,,\quad
\end{equation}
with $\Theta^* = {\mathbf{R}_{\Theta}}[g_*]\,{\cdot}\,\Theta$.
Similarly, variations of the scalar potential with respect to scalar fields can be traded for E$_{n(n)}$-variations of the embedding tensor, and this defines the extremisation conditions
\begin{equation}\label{extremization}
\,\Big(
 t_{\alpha\,M}{}^N\Theta^*_N{}^\beta
 + \Theta^*_M{}^\gamma f_{\alpha\gamma}{}^\beta \Big)
\frac{\delta V}{\delta\Theta_M{}^\beta}\big(0\,,\, \Theta^*\big)\, 
 \ =\ 0\,,
\end{equation}
as well as the scalar mass matrix by taking a second variation and projecting it onto the $\mathrm E_{n(n)}/K(\mathrm E_{n(n)})$ coset generators.
This simple idea triggered many new findings in the structure of gaugings and vacua of gauged supergravities, see for instance \cite{Dibitetto:2011gm,deRoo:2011fa,Borghese:2011en,Dibitetto:2012ia,DallAgata:2012tne,Borghese:2012qm,Kodama:2012hu,DallAgata:2012plb,Borghese:2012zs,Zwirner:2013htz,Borghese:2013dja,Catino:2013ppa,Gallerati:2014xra,Deger:2019tem,Dallagata:2021lsc}.
We can indeed attempt to directly classify solutions to the quadratic constraint \eqref{qc1} together with \eqref{extremization} with respect to an a-priori undetermined embedding tensor, and only later identify {which gauging the discovered solution belongs to}.
The above method can be further refined by looking only for solutions preserving some amount of supersymmetry~\cite{Gallerati:2014xra,Deger:2019tem}. 
In this case, one also evaluates the fermion shifts \eqref{fermionshifts} as functions of an undetermined embedding tensor, setting $\phi=0$, and imposes some BPS conditions, which now reduce to \emph{linear} constraints on the components of $\Theta_M{}^\alpha$.
In particular, using these techniques, \cite{Gallerati:2014xra} it proved that $D=4$ maximal supergravities do not admit AdS vacua with $4<\mathcal N<8$ supersymmetry, and classified all $\mathcal N\ge3$ AdS vacua.

\smallskip

An important family of new gaugings of $D=4$ maximal supergravity was constructed in \cite{DallAgata:2011aa}. 
They are generalizations of the SO(8) gauging of de{\,}Wit and Nicolai and the family of SO$(p,q)$ and CSO$(p,q,r)$ gaugings derived by Hull in the 80's \cite{Hull:1984yy,Hull:1984vg,Hull:1984rt} as analytic continuations and contractions of the SO(8) model.
We now summarize the main properties of these models.
The embedding tensor $\Theta_M{}^\alpha$ for $D=4,\ \mathcal N=8$ supergravity sits in the ${\mathbf{R}_{\Theta}}=\bf912$ representation of $E_{7(7)}$. 
The gaugings we are interested in sit within $\mathrm{SL}(8)\subset \mathrm E_{7(7)}$ and branching accordingly we find $\mathbf{56}\to\mathbf{28}+\mathbf{28}'$, $\mathbf{133}\to\mathbf{63}+\mathbf{70}$ and $\mathbf{912}\to\mathbf{36}+\mathbf{36}'+\mathbf{420}+\mathbf{420}'$, with $\mathbf{63}={\rm Adj}(\mathfrak{sl}(8))$.
The various irreps are related according to the following table
\begin{equation}
\begin{array}{|c|c|c|}\hline
  \Theta          & \mathbf{63}                          & \mathbf{70}    \\\hline
    \mathbf{28}   &     \mathbf{36}+\mathbf{420}         &  \mathbf{420}'  \\\hline
    \mathbf{28}'  &      \mathbf{36}'+\mathbf{420}'      &  \mathbf{420}
    \\\hline
\end{array}
\end{equation}
and we see that gauging a subalgebra of $\mathfrak{sl}(8)$ entails turning on only the $\mathbf{36}$ and/or $\mathbf{36}'$ components.
Denoting $A,B,\ldots = 1,\ldots,8$ indices in the fundamental representation of SL(8), one decomposes a $\bf56$ element as $V^M = (V^{AB}\,, \,V_{AB} )$ with understood antisymmetrization, while $\mathfrak{sl}(8)$ is spanned by traceless matrices $\Lambda^A{}_B$.  
We then represent the 
$\mathbf{36}$ and $\mathbf{36}'$ irreps as symmetric matrices $\theta_{AB}$ and $\xi^{AB}$, respectively, so that the non-vanishing embedding tensor components are
\begin{equation}\label{thetaxi}
\Theta_{AB}{}^C{}_D = \delta_{[A}^C\theta^{\vphantom{C}}_{B]D}\,,\qquad
\Theta^{AB,}{}^C{}_D = \delta^{[A}_D\xi_{\vphantom{C}}^{B]C}\,.
\end{equation}

The original SO(8) gauged supergravity of de{\,}Wit and Nicolai \cite{deWit:1982bul} is obtained setting \begin{equation}\theta_{AB}=g\,\delta_{AB}\qquad \text{and}\qquad \xi^{AB}=0\,,\end{equation} with $g$ a coupling constant.
Changing the signature of $\theta_{AB}$ to $p$ positive, $q$ negative and $r$ vanishing eigenvalues (with $p+q+r=8$) yields the SO$(p,q)$ models (for $r=0$) and, in general, the CSO$(p,q,r)$-gaugings.
The quadratic constraint \eqref{qc2} is automatically satisfied.

Turning on $\xi^{AB}\neq0$ leads to some surprising results~\cite{DallAgata:2011aa,DallAgata:2012mfj}.
In order to satisfy \eqref{qc2} one must impose
\begin{equation}\label{QCthetaxi}
\xi^{AB}\propto(\theta_{AB})^{-1}\qquad\text{or}\qquad\xi^{AC}\theta_{CB}=0\,,
\end{equation}
depending on whether or not $\theta_{AB}$ is invertible.
If we now set 
\begin{equation}
\theta_{AB} = g\,\cos\omega\,\delta_{AB}\,,\qquad\qquad
\xi^{AB} = g\,\sin\omega\,\delta^{AB}\,,
\end{equation}
we find a one-parameter family of `SO(8)$_\omega$' gaugings, with $\omega=0$ the original de{\,}Wit--Nicolai model.
The interpretation of the $\omega$ parameter is that of an electric-magnetic rotation of the gauge connection. 
Indeed, the vectors decompose as $A_\mu^M = (A_\mu^{AB}\,, \,A_\mu{}_{AB} )$ under SL(8), and one can characterize $A_\mu^{AB}$ as the `electric' vector fields and $A_\mu{}_{AB}$ the magnetic duals. 
This is a standard choice of symplectic frame, in which the de{\,}Wit--Nicolai theory is gauged electrically and the deformed SO$(8)_\omega$ models involve magnetic vectors in the gauge connection.
This is reminiscent of de\,Roo--Wagemans angles in half-maximal supergravity \cite{deRoo:1985jh}. In that case, an electric-magnetic rotation of a simple factor in a semisimple gauging yields physically inequivalent  theories. Here by contrast the gauge connection of a simple group is rotated altogether.
{It is very suggestive that} for any value of $\omega$, the models exhibit an SO(8) invariant, $\mathcal N=8$ supersymmetric AdS$_4$ vacuum.
The mass spectrum at this point is also $\omega$-independent.
Higher order couplings are affected by the deformation and indeed the pattern of vacua, symmetry breaking, cosmological constants and mass spectra change with $\omega$.
{
In order to classify the duality orbits of SO(8)$_\omega$ models and hence which values of $\omega$ truly yield inequivalent theories, ${\rm E}_{7(7)}$-invariant quantities were constructed out of $\Theta$ in \cite{DallAgata:2012mfj}, proving that such models are inequivalent exactly for $0\le\omega\le\pi/8$.
This range is obtained by the identifications
\begin{equation}
  \omega\simeq-\omega\,,\qquad
  \omega\simeq\omega+\frac\pi4\,,
\end{equation} found in \cite{DallAgata:2012mfj} 
of which the former is generated by a parity-related symmetry of the ungauged theory and the latter by an E$_{7(7)}$ transformation implementing an outer automorphism of the SO(8) algebra.
These identifications were reproduced explicitly through a detailed group theoretical analysis \cite{DallAgata:2014tph}, 
where it is also determined how to classify similar families of inequivalent models starting from other gauge groups.
}
\begin{figure*}[t]
  \begin{center}
    \includegraphics[scale=0.6]{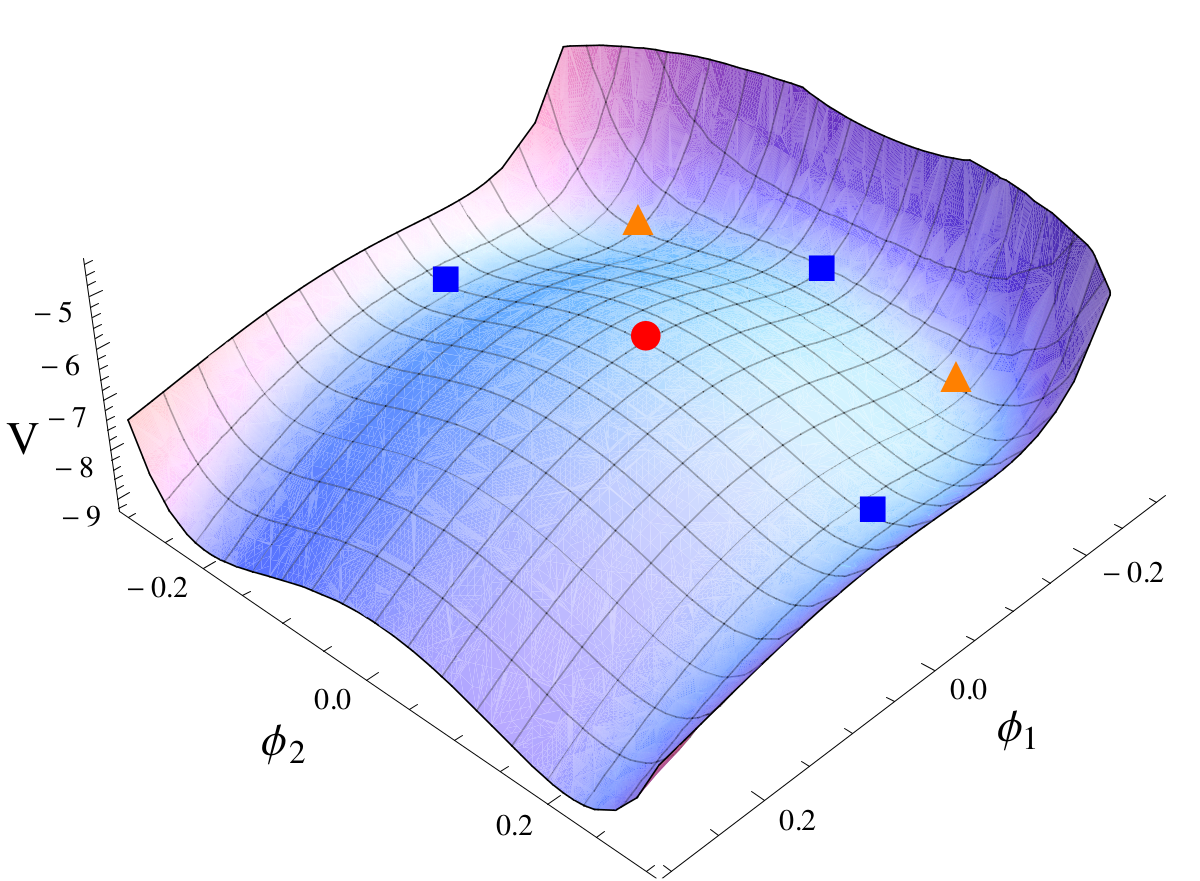} \hspace{2em}
    \includegraphics[scale=0.6]{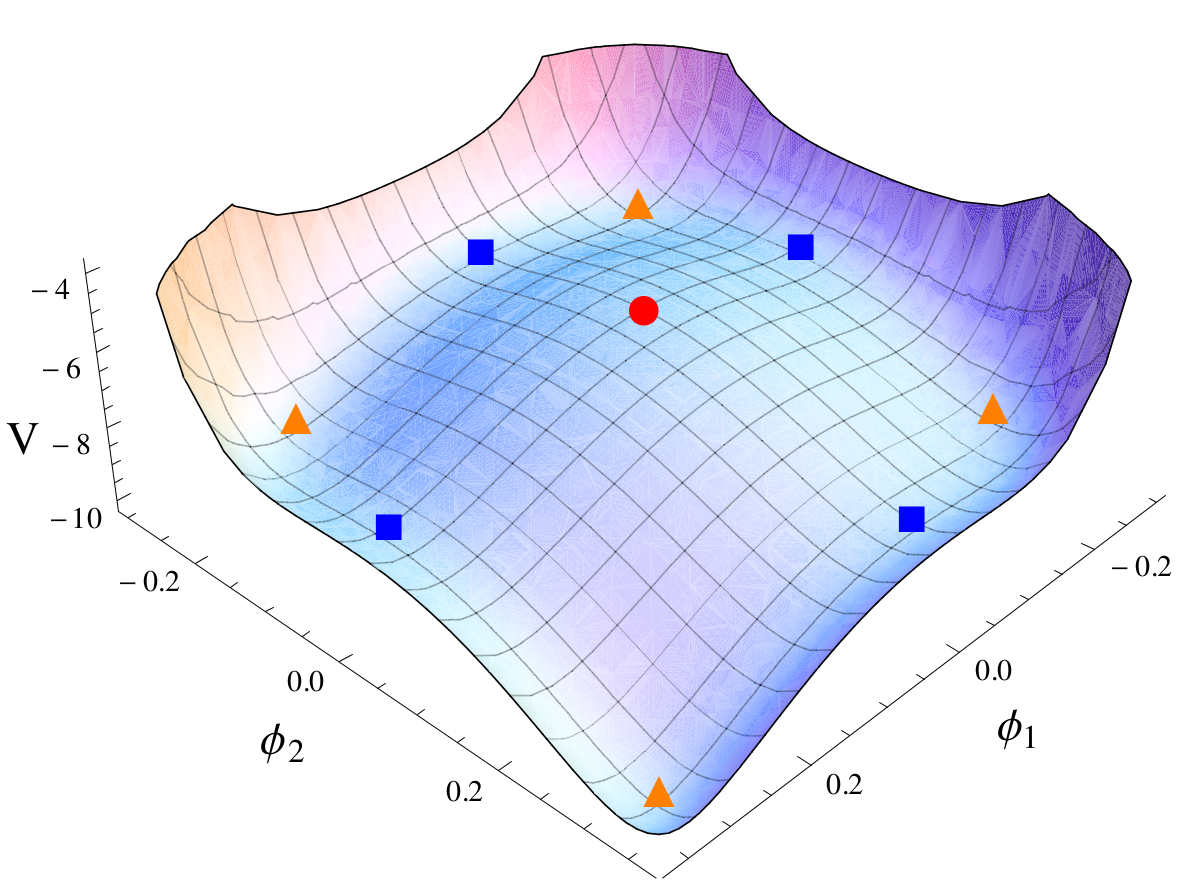}
  \end{center}
  \caption{\label{fig:vuoti}  Scalar potential of the G$_2$ invariant sector of SO(8)$_\omega$ gauged maximal supergravity for $\omega = 0$ (left) and for $\omega = \pi/8$ (right). Coloured shapes denote different vacua. Plots taken from~\cite{DallAgata:2012mfj}.}
\end{figure*}
The structure of vacua and other solutions of the SO(8)$_\omega$ models have been studied extensively both analytically and numerically \cite{Borghese:2012qm,Borghese:2012zs,Borghese:2013dja,Anabalon:2013eaa,Pang:2015mra,Karndumri:2020bkc,Berman:2022jqn}.

The physical interpretation of the $\omega$ parameter has remained elusive. 
On the one hand, it was proven in \cite{deWit:2013ija,Lee:2015xga} that the SO(8)$_\omega$  models cannot arise as consistent truncations of eleven-dimensional supergravity based on the framework of generalized Scherk--Schwarz (gSS) reductions {in EGG and ExFT, to which we come back below.} 
On the other hand, finding a possible holographic dual for the maximally supersymmetric AdS vacuum of the SO(8)$_\omega$ models is to date still an open problem.\footnote{It was initially proposed, based on the similarity of certain self-duality relations, that the $\omega=\pi/8$ theory could be dual to the ABJ theory based on gauge group $U(N)_{2}\times U(N+1)_{-2}$, but evidence of this relation is still missing.}
In a related analysis~\cite{Borghese:2014gfa}, it was found that in order to properly implement holographic boundary conditions in the $\mathcal N=8$ vacuum of the SO(8)$_\omega$  theories, supersymmetry must in fact be broken to at most $\mathcal N=3$, unless $\omega=0$.
Curiously, \cite{Borghese:2014gfa} also finds that if one truncates the SO(8)$_\omega$ models to $\mathcal N=6$, boundary conditions with $\mathcal N=6$ supersymmetry can be found for $\omega=0,\,\pi/8$. Since $\omega$ can be reabsorbed into a non-local field redefinition in $\mathcal N=6$ supergravity~\cite{Inverso:2015viq}, this result underlines a dependence on the symplectic frame in the computation of supersymmetric boundary conditions. 
To date, a systematic study of the effect of similar non-local field redefinitions on the boundary conditions of the complete $\mathcal N=8$, SO(8)$_\omega$ has not been carried out.

{\sloppy{}
While the status of the SO(8)$_\omega$ models is uncertain, other classes of gauged maximal supergravities defined by \eqref{thetaxi} have provided fertile ground for holographic applications. 
These theories have been dubbed `dyonic 
-CSO' gaugings and they involve a superposition of two CSO$(p,q,r)$ gauge groups determined by the signatures of $\theta_{AB}$ and $\xi^{AB}$ respectively, such that \mbox{$\theta_{AC}\xi^{CB}=0$} and the gauge connection is a mixture of electric and magnetic vector fields (in the natural SL(8) symplectic frame introduced above).
The gauge groups has the general form
\begin{equation}\label{dyonicCSO}
  \big(\,{\rm SO}(p,q)\times {\rm SO}(p',q') \,\big)\ltimes N
\end{equation}
}%
with $N$ a group generated by nilpotent elements of SL(8). 
The inequivalent duality orbits of such gaugings were classified in~\cite{DallAgata:2014tph}.
All these models admit one or more higher-dimensional embeddings~\cite{Guarino:2015jca,Inverso:2016eet,Inverso:2017lrz}.

The first interesting instance can be found in the ISO(7)-theories.
There are two ways to obtain an ISO(7) gauge group.
A first one~\cite{Hull:1984yy} is by setting $\xi^{AB}=0$ and choosing  $\theta_{AB}$ to have signature $(7,0,1)$.
This gauging is electric in the SL(8)-symplectic frame.
An inequivalent gauging is obtained by turning on one non-vanishing eigenvalue of $\xi^{AB}$ (so that \eqref{QCthetaxi} is satisfied).
In this case, the $\mathbb{R}^7\subset\mathrm{ISO}(7)$ subgroup is gauged by a mixture of electric and magnetic vectors.
These theories descend from consistent truncation of type IIA supergravity on $S^6$, with the non-zero eigenvalue in $\xi^{AB}$ corresponding to the Romans mass in ten dimensions~\cite{Guarino:2015jca}. 
An $\mathcal N=2$ AdS$_4$ vacuum was found to be dual to a super-CS theory with simple SU($N$) gauge group, the CS level $k$ being identified with the Romans mass.
Evidence was also given for a duality between an $\mathcal N=3$ vacuum of the theory, first found in \cite{Gallerati:2014xra}, and a super-CS theory with two adjoint chiral multiplets~\cite{Guarino:2015jca,Pang:2015rwd}.
The ISO(7) model and its duals have been studied thoroughly, see for instance~\cite{Pang:2015vna,Hosseini:2017fjo,Guarino:2020jwv}.
Among many results, it was observed that the $G_2$-invariant, non-supersymmetric vacuum of the deformed ISO(7) gauging uplifts to a perturbatively stable solution of massive IIA supergravity, also stable under certain non-perturbative decay channels~\cite{Guarino:2020jwv,Guarino:2020flh}. This would pose a challenge to the {swampland program (see for instance \cite{Palti:2019pca} for a review)} because stable, non-supersymmetric vacua are conjectured to be absent in quantum gravity. A non-perturbative decay channel was however proposed later~\cite{Bomans:2021ara}.
Another interesting application has been the construction of black holes with AdS$_4\times S^6$ asymptotics and the reproduction of their entropy through holography~\cite{Hosseini:2017fjo}, obtained by computing a topologically twisted index in the dual CFT through supersymmetric localization, along the lines of~\cite{Benini:2015eyy}.

Another interesting family of gaugings is given by models with gauge group $\big[\mathrm{SO}(6)\times X\big]\ltimes \mathbb{R}^{12}$, with $X$ equal to either $\mathrm{SO}(1,1)$, $\mathrm{SO}(2)$ or $\mathbb R$, corresponding to a hyperbolic, elliptic or parabolic generator of SL(2).
These theories arise from consistent truncations of type IIB supergravity on {${S}^5\times S^1$,%
} with an SL$(2,\mathbb{Z})$-{monodromy $\mathfrak{M}$} along the circle~\cite{Inverso:2016eet}.
{These are therefore examples of $S$-fold configurations in IIB string theory.}
The conjugacy class of the twist determines the $X$ factor of the gauge group.
It was originally observed~\cite{Inverso:2016eet} that the $\mathcal N=4$ AdS$_4$ vacuum of the $X=\mathrm{SO}(1,1)$-model, first found in \cite{Gallerati:2014xra}, {lifts to a solution of type IIB supergravity on a deformed $\tilde{S}^5\times S^1$ preserving SO(4) isometries and} can be realized as the $S^1$ compactification of (a singular limit of) one of the type IIB Janus solutions  \cite{DHoker:2007zhm}, dual to $\mathcal N=4$ super-Yang--Mills with an interface and a duality twist along $S^1$.
{While one observes that the string coupling takes finite but non-perturbative values along $S^1$ it is possible to argue that }higher-derivative corrections to the IIB supergravity action remain under control. 
{Indeed a proposal for the CFT dual of the $\mathcal N=4$ solution has been put forward in~\cite{Assel:2018vtq}. It is the IR limit
of a configuration where $\mathcal N=4$ SYM on a circle is coupled to a ${\rm T}[{\rm U}(N)]$-theory \cite{Gaiotto:2008ak}, so that the ${\rm U}(N)\times {\rm U}(N)$ flavour symmetry of the latter is gauged by the $\mathcal{N}=4$ vectors and a level–$n$ CS term is added.
The integer $n>2$ is related to the hyperbolic monodromy along $S^1$: $\mathfrak{M}=J_n\equiv \left(\begin{smallmatrix}n & 1\cr -1 & 0\end{smallmatrix}\right)\in {\rm SL}(2,\mathbb{Z})$. \par
The $\mathcal{N}=4$ vacuum is part of a three-parameter family\footnote{Recently, in \cite{Bobev:2023bxs}, a fourth non-supersymmetric deformation parameter was found.} of extrema of the potential, connected by flat directions of the scalar potential, which are expected to be dual to exactly marginal deformations of the dual SCFT. The generic point of this three-parameter manifold defines a non-supersymmetric vacuum while, in a 2-parameter subspace, the solution is $\mathcal{N}=2$ with ${\rm U}(1)_F\times {\rm U}(1)_R$-symmetry ($\mathcal{N}=2\& {\rm U}(1)^2$ in short) and, at a single point, $\mathcal{N}=4$. {The surface of $\mathcal{N}=2$ vacua features a special point at which the symmetry is enhanced to ${\rm U}(2)$.} The corresponding solution in $D=10$ was constructed starting from the $D=4$ gauged model in \cite{Guarino:2020gfe} and from ${\rm SO}(6)$-gauged $D=5$ model in \cite{Bobev:2020fon}. 
Two of the flat directions of the scalar potential (whose parameters are typically denoted by $\chi_i$) were interpreted as compact metric moduli of the internal manifold \cite{Giambrone:2021zvp}, induced by Wilson lines along $S^1$ of gauge vectors in the CSS compactification of the $D=5$ ${\rm SO}(6)$-gauged model \cite{Guarino:2021hrc,Berman:2021ynm}. The third modulus, which connects the $\mathcal{N}=2\&{\rm U}(2)$ solution to the $\mathcal{N}=4$ one through a line of $\mathcal{N}=2\& {\rm U}(1)^2$ vacua, has a non-trivial geometric characterization. 
The corresponding solutions were studied, also from the holographic point of view, in \cite{Arav:2021gra,Bobev:2021yya,Bobev:2023bxs} and the Kaluza-Klein spectrum on them was computed in \cite{Cesaro:2021tna}. In \cite{Giambrone:2021wsm}, evidence was given that the non-supersymmetric Type IIB solutions obtained through $\chi_i$-deformations of the $\mathcal{N}=4$ S-fold, are stable. In particular, perturbative stability was proven through the computation of the Kaluza-Klein spectrum on such backgrounds, which exhibits no scalar masses below the Breitenlohner-Freedman (BF) bound \cite{Breitenlohner:1982jf}. If proven  non-perturbatively stable, such solutions would provide, in the dual holographic picture, the first instance of a non-supersymmetric conformal manifold, besides being in tension, on the bulk theory side, with predictions from the swampland program \cite{Palti:2019pca}, as mentioned earlier.
Supersymmetric domain wall and black hole solutions asymptoting the $\mathcal{N}=4$ and $\mathcal{N}=2$ S-fold backgrounds were constructed and studied in 
\cite{Guarino:2021kyp,Bobev:2023bxs,Guarino:2024zgq}.
The $X={\rm SO}(1,1)$ theory also features two other families of marginally connected vacua: a three-parameter manifold of (unstable) $\mathcal{N}=0$ \& ${\rm U}(1)^3$ vacua and a two-parameter one of $\mathcal{N}=1$ \& ${\rm U}(1)^2$\cite{Guarino:2019oct}. The former contains a maximally symmetric point with symmetry ${\rm SO}(6)$, which uplifts to a ${\rm AdS}_4\times S^1\times S^5$  S-fold. In the latter class, on the other hand, the symmetry is enhanced, at a point, to ${\rm SU}(3)$ and the corresponding Type IIB solution has geometry ${\rm AdS}_4\times S^1\times \tilde{S}^5$, where $\tilde{S}^5$ is locally described as $\mathbb{CP}^2\times S^1$. These solutions were generalized in \cite{Bobev:2019jbi} to backgrounds of the form ${\rm AdS}_4\times S^1\times M^5$, where $M^5$ is a Sasaki-Einstein space. The non-supersymmetric ${\rm AdS}_4\times S^1\times S^5$ S-fold suggests 
a general paradigm for the construction of ${\rm AdS}_{d-1}\times S^1\times S^d$ U-folds~\cite{Astesiano:2024gzy} which was applied, in the same work, to the construction of the ${\rm AdS}_{2}\times S^1\times S^3\times CY_2$  solutions to Type IIB supergravity, where $CY_2$ can either be a 4-torus $T^4$ or $K3$, starting from the near-horizon geometry of a system of D1-D5 or of F1-NS5 branes in which the five branes are wrapped on $CY_2$.\footnote{The monodromy $\mathfrak{M}$ is either in ${\rm O}(4,4;\mathbb{Z})$ or in ${\rm O}(4,20;\mathbb{Z})$, depending on the two possible choices for $CY_2$.}
}\par
While most of the focus has been on hyperbolic duality twists, where more supersymmetric AdS solutions have been found, the elliptic case (with $X=\mathrm{SO}(2)$) also exhibits supersymmetric solutions~\cite{Berman:2021ynm}, and in this case one may also construct globally geometric solutions by appropriately choosing the $S^1$ periodicity.

As another example of the richness of the models in \eqref{thetaxi}, a very large class of gauged supergravities exhibiting Minkowski vacua with various degrees of supersymmetry breaking was discovered starting from the $\omega=\pi/4$ SO(6,2) gauging~\cite{Catino:2013ppa}.
This turns out to be equivalent to the SO*(8)-gauging by Hull \cite{Hull:2002cv} and has a non-susy Minkowski vacuum with the same mass spectra as a CSS gauging. 
Through a procedure of singular limits in moduli space, infinite families of gaugings can be constructed with guaranteed Minkowski vacua.
The mass spectra and susy breaking patterns are entirely determined by charge assignements under a U$(1)^4$ symmetry, generalising the structure of Cremmer--Scherk--Schwarz gaugings. This has made it possible to compute the supertraces of the mass spectra and hence one-loop corrections to the scalar potential \cite{DallAgata:2012tne}.
Double copy constructions for amplitude computations in such models have also been investigated \cite{Chiodaroli:2018dbu}.

\section{Consistent KK truncations and new gaugings}

Until recently, it was unclear how to systematically characterize and construct consistent Kaluza--Klein truncations from ten- and eleven-dimensional supergravity to gauged (maximal) supergravities.
Significant progress has been made in the context of generalised Scherk--Schwarz reductions in EGG and ExFT.
We now give a few more details on these frameworks (see also \cite{Duff:2025tot} in this volume) and describe many of the recent results in this line of work.
A gSS truncation ansatz in $\mathrm E_{n(n)}$ ExFT is determined by a `twist matrix' or `generalized frame' $E_{\dot M}{}^M(y)$ taking values in E$_{n(n)}\times\mathbb R^+$ and dependent on the internal space coordinates $y^a$.\footnote{The standard SS reductions correspond to a frame equal to the left-invariant forms on a group manifold $G$ (embedded in $\mathrm E_{n(n)}\times\mathbb R^+$ as elements of a GL$(d)$ subgroup), while in the CSS setup the frame depends on a circle and takes values in $\mathrm E_{n-1(n-1)}\subset E_{n(n)}$.}
Here we temporarily use dotted indices to distinguish where the factorisation has been carried out so that objects with undotted indices belong to ExFT and those with dotted indices belong to the $D$-dimensional gauged maximal supergravity obtained after reduction.
The generalized frame must satisfy a first-order PDE of the schematic form
\begin{equation}\label{gSS}
\mathcal{L}_{E_{\dot{M}}} E_{\dot{N}}{}^M = -X_{\dot M \dot N}{}^{\dot P} E_{\dot P}{}^M\,,
\end{equation}
based on a generalized notion of Lie derivative $\mathcal L$ along the internal manifold, which we refrain from displaying explicitly here, and which encodes the action of internal infinitesimal coordinate transformations and $p$-form gauge symmetries on the ExFT field content~\cite{Berman:2012vc,Coimbra:2011ky}.
Here $X_{\dot M\dot N}{}^{\dot P}$ must be constant and determines the embedding tensor of the gauged maximal supergravity obtained from the truncation (for Lagrangian gaugings, this is simply $X_{\dot M\dot N}{}^{\dot P} = \Theta_{\!\dot M}{}^\alpha t_{\alpha \dot N}{}^{\dot P}$)
The framework can be extended to include consistent truncations of massive IIA supergravity~\cite{Ciceri:2016dmd}.

The framework above allowed to prove no-go theorems for the uplift of the $\omega$-deformed SO(8) models~\cite{deWit:2013ija,Lee:2015xga}, as well as for their non-compact SO($p,q$)$_\omega$ siblings~\cite{Inverso:2017lrz}. 
The standard SO and CSO gaugings in various dimensions could be systematically uplifted on spheres and hyperboloids, depending on the signature~\cite{Lee:2014mla,Hohm:2014qga}, and the dyonic CSO gaugings \eqref{dyonicCSO} can also be systematically uplifted \cite{Inverso:2016eet}.
Among many examples, consistent Pauli reductions could be realised~\cite{Baguet:2015iou} where the full $G_L\times G_R$ isometries of a group manifold $G$ is reflected as gauging upon truncation.
As another significant example, the full four-parameter family of Cremmer--Scherk--Schwarz gaugings~\cite{Cremmer:1979uq}, realising supersymmetry breaking in Minkowski space, can be uplifted geometrically to eleven dimensions~\cite{DallAgata:2019klf}. Such gaugings are parametrised by four masses, of which three realize the original Scherk--Schwarz setup~\cite{Scherk:1979zr} on a twisted torus, while the last one requires EGG or ExFT to be properly uplifted to eleven dimensions.
Further models have been recently constructed in different dimensions. 
For instance, several consistent truncations on AdS$_3\times S^3\times T^4$ and AdS$_3\times S^3 \times S^3 \times S^1$ to (half-)maximal gauged supergravities have been constructed, which have allowed to identify large spaces of moduli of these solutions~\cite{Eloy:2021fhc,Eloy:2023zzh,Eloy:2024lwn}.
The framework of generalised Scherk--Schwarz reductions has by now been developed for truncations down to $D\geq2$ gauged maximal supergravities~\cite{Bossard:2023wgg}.

Since it is established that only a subset of all gauged (maximal) supergravities can arise from consistent Kaluza--Klein truncations, it becomes important to determine under what conditions a given gauging admits an uplift to ten or eleven dimensions and in case, what is the geometry of the internal space and explicitly construct the reduction ansatz.
The requirement for the existence of an uplift (based on the gSS ansatz) can be cast in a duality invariant way, by constructing a coset space $G_g/H$ from the gauge group $G_g$ and imposing some algebraic constraints on the associated embedding tensor.
One then requires that the matrix $\Theta_M{}^{\underline m}$, in which the adjoint index has been projected to a set of coset generators, solves the \emph{section constraint}\footnote{We have now dropped the dots on gauged supergravity indices.}
$
Y^{MN}{}_{PQ}\,\Theta_M{}^{\underline m} \Theta_N{}^{\underline n} \ =\ 0
$, where
$Y^{MN}{}_{PQ}$ is an E$_{n(n)}\times \mathbb R^+$ invariant tensor known from ExFT, which entirely characterizes the $\mathcal L$ operator appearing in \eqref{gSS}~\cite{Berman:2012vc}.
A second constraint (expressed here for Lagrangian gaugings) requires the combination $\Theta_P{}^{\underline m}t_{\underline m\,M}{}^P$ to be on section as well.
The gSS frame is then explicitly constructed from geometric data on $G_g/H$ and by computing and integrating certain background fluxes, which can be done explicitly and systematically.
We refer to~\cite{Inverso:2017lrz} for more details (see \cite{duBosque:2017dfc} for earlier results).
These results can be rephrased and refined in the language of Lie algebroids and their generalizations \cite{Bugden:2021wxg} (see also \cite{Hassler:2022egz} for a different reinterpretation of such results).
The extension to $D=3$ gauged supergravity was carried out only recently, due to technical complications in E$_{8(8)}$ exceptional geometry \cite{Inverso:2024xok}.
Duality-invariant \emph{necessary} conditions for an uplift to exist, which do not require scanning through choices of $G_g/H$ and are therefore useful to derive no-go results,  have also been identified for $D=3,2$ in~\cite{Eloy:2023zzh,Bossard:2023wgg,Inverso:2024xok}.
For instance, in $D=3$ one has ${\mathbf{R}_{\Theta}}={\bf1}+{\bf3875}$ but the singlet must vanish for a gSS uplift to exist.

One can alternatively break E$_{n(n)}$ invariance to GL$(d)$ ($d$ being the dimension of the internal space) and phrase the uplift conditions as a restriction on the GL$(d)$ irrep content of the embedding tensor~\cite{InversoTalk}. This idea has now been applied in all dimensions down to $D=2$ \cite{Hassler:2022egz,Bossard:2023jid,Inverso:2024xok}. 
Such linear conditions can be a promising starting point for classification efforts. %
They can also be naturally combined with the algebraic search for vacuum solutions described in the previous section, giving a powerful approach to solution generation that has not yet been put to use.

\section{Gaugings in two dimensions}

Outliers in these analyses are $D=2$ gauged maximal supergravities.
Two-dimensional gauged supergravities are the realm of (near) AdS$_2$ solutions and hence of near-horizon geometries of black holes. 
Models such as JT (super)gravity (see for instance~\cite{Mertens:2022irh} and references therein) may also be obtained as truncations or perturbations around a vacuum of some $D=2$ gauged supergravity, the latter possibly arising itself from ten or eleven dimensions through a consistent Kaluza--Klein truncation.
Two-dimensional (super)gravities with scalars parametrising a symmetric space are classically integrable and this is reflected in the infinite-dimensionality of their Geroch-like global symmetry group \cite{Geroch:1970nt}, which for maximal supersymmetry is the affine Kac--Moody group E$_9$~\cite{Julia:1981wc}.
Integrability can be encoded in a linear system -- foregoing duality covariance -- or as a generalization of a covariant twisted self-duality constraint, in analogy with the general relation~\eqref{tsd}.
The latter approach~\cite{Julia:1996nu,Bossard:2021jix} was crucial in the recent progress on $D=2$ gaugings and the construction of E$_9$ exceptional field theory~\cite{Bossard:2017aae,Bossard:2018utw,Bossard:2021jix}. 

In contrast with higher-dimensional models, a systematic construction of $D=2$ gauged supergravity Lagrangians has long been beyond reach, because of the complicated representation theory of the maximal unitary subgroup $K(E_9)$ of E$_9$, which acts on fermions.
Branching ${\mathbf{R}_{\Theta}}$ to construct fermion shifts, Yukawa couplings and the scalar potential has so far been impossible (see~\cite{Kleinschmidt:2021agj} for recent progress).
Nonetheless, much better control over the bosonic sector of $D=2$ gauged supergravities has recently been achieved.
The structure of the bosonic sector of $D=2$ gauged supergravities was first developed in \cite{Samtleben:2007an}, but the scalar potential could not be constructed.
The only complete and supersymmetric construction has been achieved (bypassing a duality covariant formulation) for SO(9) gauged supergravity (and its analytic continuations) \cite{Ortiz:2012ib,Anabalon:2013zka}.
This model arises from consistent truncation of IIA supergravity on $S^8$ and includes a 1/2 BPS solution lifting to the near-horizon limit of the D0 brane~\cite{Itzhaki:1998dd} and is therefore relevant for the holographic study of the regularized supermembrane (or BFSS) matrix model~\cite{deWit:1988wri}.
The proof of consistent truncation was only carried out recently with the development of E$_9$ exceptional field theory~\cite{Bossard:2022wvi,Bossard:2023jid}.
This has also allowed to construct a fully duality covariant formalism for the bosonic sector of $D=2$ gauged maximal supergravity and to identify the scalar potential of all models that admit a higher dimensional origin, thus capturing the full bosonic dynamics.
In $D=2$ , the embedding tensor is infinite-dimensional.
Nonetheless, only a finite (but very large) set of coefficients descends from consistent Kaluza--Klein truncations and the physical Lagrangian always reduces to a finite expression.
The potential takes a rather simple covariant form,
with $\theta_M$ the embedding tensor in the basic E$_9$ representation, $[\eta_{-2}]{}^M{}_P{}^N{}_Q$ the level 2 coset Virasoro generator and $\rho$ the dilaton:
\begin{equation}
2\rho^3\,V = \theta_M\theta_N\,\mathcal{M}^{MN} + \rho^2\, [\eta_{-2}]{}^M{}_P{}^N{}_Q\,\theta_M\theta_N\,\mathcal{M}^{PQ}\,.
\end{equation}


\section{Outlook}

As stressed already, no full classifications of gauged (maximal) supergravities exist, nor do the subset of such models descending from consistent KK truncation from ten or eleven dimensions. 
The tools summarized in this contribution do at least render such an endeavour more feasible.
One must also be careful since the uplift to a gauged supergravity may lead to a non-compact internal space.
Compactness may sometimes only be achieved by discrete quotienting, or not at all.
Quotienting is under control for SS reductions on group manifolds but less explored for gSS reductions, especially for locally geometric identifications, such as the ones described for $S$-fold solutions.
Finding necessary and sufficient conditions for the compactness of the internal manifold is therefore an important problem.
Some conditions, algebraic in $\Theta$, were presented recently \cite{Inverso:2024xok}, but a systematic analysis is missing.

There are also many reasons to go beyond maximal supersymmetry and look at gauged supergravities with fewer supercharges.
Consistent KK truncations preserving only a fraction of the original supercharges are best encoded in terms of generalized $G$-structures in EGG and ExFT~\cite{Cassani:2019vcl}. 
Partial classifications of gauged models that can admit an uplift based on such framework have been carried out~\cite{Josse:2021put}.
It is an interesting and open question to determine whether and to what extent one could reduce the conditions for the existence of an uplift to algebraic ones on the gauged supergravity data, and explicitly construct the internal manifold and reduction data in analogy with the gSS case described earlier.
Another surprising outcome of the framework of generalized $G$-structures has been the identification of an unexpected way to carry out a consistent truncation from $\mathcal N=8$ to $\mathcal N=4$ supergravity in $D=4$~\cite{Guarino:2024gke}, which deserves to be studied further and generalized to other examples.

The structure of $D=2$ gauged (maximal) supergravities and their higher-dimensional origins have proven very complex and rich at the same time.
A full analysis of uplift conditions for $D=2$ gauged maximal can be reasonably expected to work along the same lines as their higher dimensional counterparts and an explicit classification of couplings potentially admitting uplift is already available~\cite{Bossard:2023jid}.
The tools are therefore ready for exploring the vast landscape of $D=2$ gauged maximal supergravities, of their vacuum solutions (using the same homogeneity trick described in \eqref{gtto}) and of their uplift geometries.
Still, it would be desirable to gain control over supersymmetry in $D=2$ in a systematic way, overcoming the difficulties in dealing with the representation theory of $K(\mathrm E_9)$. This would be especially important in the construction of BPS solutions from $D=2$ gauged supergravities and would also open the door to the study of KK reductions based on generalised $G$ structures.

We have described how gauged supergravities arise as consistent subsectors of ten- or eleven-dimensional supergravities on certain backgrounds. In the context of holography, the fact that a gauged supergravity describes a consistent truncation of a ten- or eleven-dimensional maximal theory implies the existence of a consistent subspace of operators in the dual CFT, at least in an appropriate large $N$ limit. Although highly desirable, a rigorous characterization of this basic idea on the field theory side of teh correspondence is still lacking.
It may also help clarify whether gauged supergravities with no geometric gSS uplift can still have a valid  physical interpretation beyond the classical limit, and possibly, a string theory origin.
  
Indeed, considering the problem of uplifting a gauged supergravity, an important question to explore is whether it can arise entirely non-geometrically.
We have already described how gauged supergravities associated with non-geometric backgrounds such as S-folds can give rise to consistent string theory solutions.
One may ask whether by looking beyond the search for uplifts to ten- and eleven-dimensional supergravities and working directly in string theory, one may find a stringy interpretation for gaugings without a higher-dimensional embedding captured by generalized geometry.
A notion of non-geometry is available in double and exceptional field theories, based on a violation of their so-called `section constraint', and has been extensively used to give an interpretation to gauged supergravities without a geometric uplift, including the SO(8)$_\omega$ models~\cite{Lee:2015xga}.
The stringy interpretation of such constructions is somewhat clear in the case of gaugings arising from twisted double tori, which can often be viewed as asymmetric orbifolds (without fixed points) on the string worldsheet, see for instance~\cite{Hellerman:2002ax,Hull:2003kr,DallAgata:2019klf}.
However, in all other cases where some form of curvature in the internal space is expected, the connection between non-geometry in double and exceptional field theory and non-geometry in string theory has not been clarified.
A better understanding of these issues, presumably requiring to work directly in string theory rather than supergravity and generalised geometry, would be highly desirable.

\section*{Acknowledgements}

The authors wish to thank Anna Ceresole and Gianguido Dall’Agata for the kind invitation to contribute to this volume.
We would also like to thank our collaborators and in particular
G. Bossard,
F. Ciceri,
G. Dall'Agata, 
A. Giambrone, 
A. Guarino, 
A. Kleinschmidt,
E. Malek, 
H. Samtleben, 
C. Sterckx, 
B. de Wit,  
with whom several of the results discussed in the present review were obtained.

\begingroup\raggedright\endgroup

\end{document}